# First-principles investigation of thermodynamics and electronic transitions in vacancy-ordered rare-earth perovskite nickelates

Devang Bhagat, Ranga Teja Pidathala and Badri Narayanan[*]

Department of Mechanical Engineering, University of Louisville, Louisville KY 40292, USA

## ABSTRACT

Controlled introduction of oxygen vacancies offers an effective route to induce metal-to-insulator transition in strongly correlated rare-earth nickelates ($RNiO_3$) at room temperature. However, the role played by the rare-earth cations on the structure, thermodynamic stability, and electronic properties of oxygen-deficient nickelates remains unclear. Here, we employ density functional theory calculations with Hubbard corrections (DFT + $U$) to investigate the whole family of $RNiO_{2.5}$ ($R$ = Pr – Er) compounds in two commonly observed oxygen-vacancy ordered configurations, namely brownmillerite, and square planar. We find that square planar polymorph is always more stable (~0.4 eV/u.f) than the brownmillerite for all rare-earth cations, owing to the exceedingly low volumetric strains (< 1%). Interestingly, the formation energy of $RNiO_{2.5}$ gradually increases with decreasing size of rare-earth cation owing to stronger Ni-O covalent interactions in pristine $RNiO_3$ with small $R$. This necessitates more oxygen-lean environments for synthesis of $RNiO_{2.5}$ with smaller rare-earth cations. Analysis of the density of states and band structures reveals that electronic structure of $RNiO_{2.5}$ is governed by two factors: (a) localization of electron on $NiO_6$ octahedra yielding a Mott insulating state with strong correlations as Ni $e_g$ is half filled, (b) crystal field splitting in the $NiO_4$ tetrahedra/square planar configuration. Brownmillerite $RNiO_{2.5}$ is metallic, while square planar $RNiO_{2.5}$ is an insulator with a predicted gap of ~0.2 – 0.3 eV, depending on the rare-earth cation. Crystal orbital Hamilton population (COHP) analysis indicates that the Ni-O bond belonging to square-planar $NiO_4$ configurations exhibit much greater covalent character than $NiO_6$ octahedra in square planar $RNiO_{2.5}$. These findings will serve as a guide for future synthesis efforts on oxygen-deficient rare-earth nickelates, and design of quantum materials for next generation neuromorphic computing architectures.

---

[*]Corresponding author, Email: badri.narayanan@louisville.edu

## I. INTRODUCTION

The strong interplay between lattice, electronic, and magnetic degrees of freedom in complex oxides gives rise to numerous exotic functionalities, including superconductivity, magnetism, charge/spin ordering, multiferroicity, superionic conduction, and catalysis.[1-7] Rare-earth nickelates ($R$NiO$_3$; $R$ = Pr–Er) with a perovskite (ABO$_3$; R$^{3+}$ at A-site, and Ni$^{3+}$ at B-site) crystal structure are a prototypical family of strongly correlated materials that offer extraordinary promise for use in synaptic transistors,[8-11] tunable photonics,[12] electric field sensors,[13] superionic lithium conductors,[4] bio-electronic interfaces,[14] and several other technologies.[8, 15, 16] This promise is primarily fueled by the ability of $R$NiO$_3$ to undergo reversible metal-to-insulator transition (MIT) under external stimuli (e.g., temperature), often accompanied by changes in magnetic properties. Importantly, MIT in $R$NiO$_3$ can be effectively tuned by controlling the covalence of Ni-O bonds[6, 17] using chemical or hydrostatic pressure, which opens doors to engineer novel electronic phases via strain confinement.[7, 18]

Thermal activation is the most typical approach to induce MIT in perovskite nickelates. Upon cooling, all $R$NiO$_3$ ($R$ = Pr–Er) undergo MIT at a certain temperature $T_{MI}$ with a concomitant structural transition from an orthorhombic (*Pbnm*) to monoclinic (*P2$_1$/n*) phase, suggesting the presence of a strong lattice-electron coupling in these materials.[1, 19, 20] Previous studies have characterized the metallic ($T > T_{MI}$) and insulating ($T < T_{MI}$) phases by acknowledging the mixed ionic-covalent character of $R$NiO$_3$.[21, 22] Essentially, in the metallic phase (above $T_{MI}$), all Ni atoms adopt an electronic configuration of *3d$^8$L* (*L*: ligand hole on surrounding O atoms) rather than *3d$^7$* as expected for formal valence of 3+ on Ni in $R$NiO$_3$. At $T < T_{MI}$, the symmetry-lowering structural transition introduces breathing distortions in the lattice, and causes transfer of ligand holes between neighboring Ni due to covalency of Ni-O bond, yielding two sets of Ni atoms with distinct electronic configurations, namely: 1) *3d$^8$* , and 2) *3d$^8$L$^2$*. From a technological perspective, it is important to note that the transition temperature $T_{MI}$ can be effectively tuned by modulating Ni-O covalency (i.e., energy required to transfer ligand holes) by tuning the composition of rare-earth cations[23] or applied strain.[1, 24] Furthermore, $T_{MI}$ can also be tailored by using electric field,[25] and isotope substitution.[26] Nevertheless, thermally activated MIT is not practical for emerging electronic devices (e.g., resistive random access memory devices) due to their reliance on precise temperature control and low switching speeds.

Recently, electron-doping of $R$NiO$_3$ at high concentrations by using intercalating ions (e.g., H$^+$, Li$^+$) or fluorine substitution of oxygen has emerged as a lucrative route to achieve colossal changes in resistivity (~6 orders of magnitude) under ambient conditions.[4, 8, 9, 13, 27] In these cases, the added electron localizes on the NiO$_6$ octahedra, transforming the electronic configuration from $3d^8\underline{L}$ to $3d^8$ configuration (i.e., $t_{2g}^6 e_g^2$); the on-site Coulomb repulsion in $e_g$ opens up a large Mott-gap.[28] An alternative approach to achieve electron-doping of $R$NiO$_3$ is to introduce oxygen vacancies in a controlled manner. Indeed, Kotiuga *et al.* recently employed low oxygen pressure environment using a Mg trap to systematically control oxygen vacancy (OV) concentration in SmNiO$_3$ thin films to achieve dramatic changes in resistivity (several orders) at room temperature.[29] Importantly, OVs can be redistributed by using electric fields to obtain a memristive behavior. Other experimental works have also reported increase in resistivity in oxygen-deficient NdNiO$_3$ and LaNiO$_3$ films.[11, 30] Similar to the electron-doping mediated by ions, the electron coming from OVs also localize on the NiO$_6$ octahedra resulting in a Mott state. However, introduction of OVs causes changes in Ni-O coordination, resulting in a wide range of possible NiO$_x$ polyhedra (e.g., NiO$_5$ pyramids, NiO$_4$ tetrahedra, and NiO$_4$ square planar) – each with a distinct crystal field splitting of Ni $3d$ orbitals depending on the local geometry of the O ligands. Unlike the case of intercalating ions, the crystal field splitting of the various polyhedral shapes in the oxygen-deficient nickelate would impact the overall electronic structure, and the consequent band gap.[29] Experiments also show that exceptional changes in resistivity mostly occur at high concentration of OVs, such as that corresponding to a stoichiometry of $R$NiO$_{2.5}$ (which amounts to addition of 1$e^-$/Ni).[29] At such OV concentrations, OVs would typically arrange in a spatially ordered manner, yielding a variety of vacancy-ordered crystal structures. For instance, in oxygen-starved environment, ferrites and cobaltates are known to undergo topotactic transition from perovskite to a brownmillerite phase featuring a ordered arrangement of NiO$_6$ octahedra and NiO$_4$ tetrahedra; this phase transition can be reversed upon annealing in O-rich environment.[31, 32]

Notwithstanding these advances, until now, studies on MIT triggered by electron-filling Mott transition in $R$NiO$_3$ by OVs is limited only to a few works on SmNiO$_3$ and NdNiO$_3$.[11, 29, 30] This leaves several open questions related to the effect of rare earth cation on (a) thermodynamic stability, (b) MIT induced by OVs, and (c) modulation of electron gap in oxygen-deficient $R$NiO$_3$, especially at high OV concentration (where vacancy-ordered phases are observed). Note, the size of rare-earth cation is well known to impact the covalency of Ni-O bonds in $R$NiO$_3$;[17] which, in turn, is expected to impact the structure, thermodynamics, and electronic properties of these materials owing to the strong lattice-electron coupling. Here, using $R$NiO$_{2.5}$ as a representative stoichiometry for high OV concentration, we employ density functional theory calculations (DFT + $U$) to systematically investigate the effect of rare earth cations on structure, energetics, Ni-O bonding character, and electronic properties for two commonly observed vacancy-ordered configurations, namely, brownmillerite (consisting of alternating planes of NiO$_6$ octahedra and NiO$_4$ tetrahedra) and square planar (with chains of corner-shared NiO$_4$ square planar polyhedra and NiO$_6$ octahedra). [33-35]

## II. METHODS

### A. Crystal structures

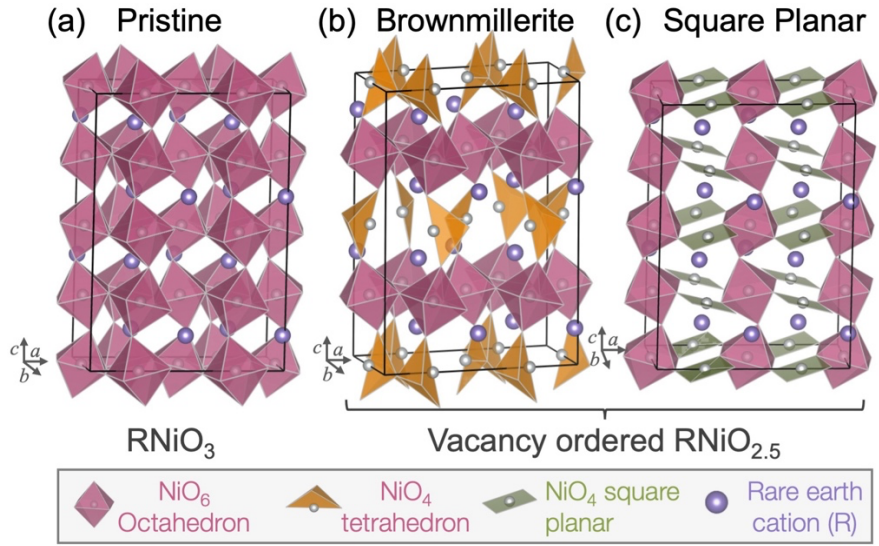

**Figure 1. Crystal structure of pristine and vacancy-ordered rare-earth nickelates.** Schematic illustration of atomic structure of (a) pristine $R$NiO$_3$, and two polymorphs of vacancy-ordered $R$NiO$_{2.5}$, namely (b) brownmillerite, and (c) square-planar. The NiO$_6$ octahedra, NiO$_4$ tetrahedra, and NiO$_4$ square planar configurations are shown in red, tan, and colors, respectively. The rare-earth cations are depicted as purple spheres.

Rare-earth nickelates ($R$NiO$_3$; $R$ = Pr–Er) exhibit a distorted perovskite structure, which consists of a 3-D network of corner-sharing NiO$_6$ octahedra, and rare-earth ($R$) cations occupying the interstitial spaces for electrical neutrality (Figure 1(a)).[1, 19] Note, an ideal perovskite structure is cubic; however, the small size of $R$ cations introduces distortions in the ideal cubic lattice by facilitating tilting of NiO$_6$ octahedra. Depending on the extent of distortions, two distinct polymorphs are possible for $R$NiO$_3$, namely, orthorhombic and monoclinic, with space groups *Pbnm* and ($P2_1/n$), respectively. [1, 17, 19, 20] The orthorhombic phase typically arises due to the rotation of NiO$_6$ octahedra around the [110] and [001] directions in ideal cubic perovskite.[19] On the other hand, the monoclinic structure is characterized by antiferrodistortive octahedral rotations, breathing of NiO$_6$ octahedra, antipolar motion involving rare-earth cations, Jahn-Teller distortions, and antiferroelectric motion.[17]

Previous experiments have shown that metal-insulator transition can be initiated in $R$NiO$_3$ by introducing 1e$^-$/Ni, either using chemicals (e.g., hydrogen, lithium)[4, 9] or via OVs.[29] In terms of OVs, the critical stoichiometry at which insulating phase appears has been reported to be $R$NiO$_{2.5}$, which corresponds to OV concentration of ~16.6 at % (causing doping of 1e$^-$/Ni).[29] At this stoichiometry, two polymorphs -- each with different spatial ordering of OVs -- have been previously synthesized for other transition metal oxides. These polymorphs include: 1) Brownmillerite, with alternating layers of NiO$_4$ tetrahedra and NiO$_6$ octahedra along the crystallographic *c* axis (Figure 1(b)), which have been observed for cobaltites[33] and ferrites;[34] and 2) Square Planar, with chains of corner-sharing NiO$_6$ octahedra and NiO$_4$ square planar configurations (Figure 1(c)), as reported in LaNiO$_{2.5}$.[35] We have employed both these polymorphs to study the effect of rare-earth cation on structure, energetics, and electronic structure of $R$NiO$_{2.5}$.

**B. Computational Details**

We performed spin-polarized density functional theory (DFT + $U$) calculations with Hubbard corrections as implemented in Vienna *Ab initio* Simulation Package (VASP).[36, 37] All the DFT + $U$ calculations are performed within the framework of generalized gradient approximation (GGA) using the projector augmented wave formalism to treat the interactions between valence electrons and ions.[38] Exchange correlations are described using the Perdew-Burke-Ernzerhof functional specifically designed for solids (PBEsol)[39] using pseudopotentials: Ni_pv (valence: $3p^64s^23d^8$), O (valence $2s^22p^4$) and Sm_3 (valence: $5s^25p^26s^24f^1$) supplied by VASP.[36, 37] The remaining five f-electrons in Sm are frozen to avoid self-interaction errors.[40] For all the other rare-earth cations, we froze all the f-electrons, while treating (a) the *5s*, *6s*, *5p*, and *5d* in Pr and Nd, and (b) *6s*, *5p*, and *5d* electrons Eu, Gd, Tb, Dy, Yb as valence electrons.[40] Hubbard correction with $U$ = 2 eV is used to treat electron localization for Ni atoms. Note that previous works have reported that PBEsol + $U$ (with $U$ = 2 eV) using the above mentioned pseudopotentials correctly predicts the ground state for the entire series of pristine $R$NiO$_3$ with band gaps in excellent accordance with experiments.[17,

[41, 42] Additionally, this approach has also been reported to correctly capture MIT in $R$NiO$_3$ induced by electron doping via protons,[43, 44] lithium,[45] aluminum, and beryllium ions.[46]

**Table 1. Structural parameters of pristine $R$NiO$_3$.** Optimized structural parameters of monoclinic polymorph of $R$NiO$_3$ with FM ordering obtained from DFT + $U$ calculations. The lengths of unit cell vectors, namely $a$, $b$, and $c$, volume $V$, monoclinic angle $\beta$, and inter-octahedral Ni–O–Ni tilt angle in a standard unit cell of monoclinic $R$NiO$_3$ are provided for nine different rare-earth cations $R$ = Pr–Er with varying ionic radii, $r_R$ = 0.99–0.89 Å. The difference in cohesive energy of monoclinic (M) and orthorhombic (O) phases (i.e., $\Delta E_{M-O}$) are listed. Deviation of DFT-predicted value from experiments (wherever available) are provided within parenthesis.

| $R$ | $r_R$* (Å) | $a$ (Å) | $b$ (Å) | $c$ (Å) | $\beta$ (deg.) | Volume (Å³/u.f.) | Ni-O-Ni (deg) | $\Delta E_{M-O}$ (meV/u.f) |
|---|---|---|---|---|---|---|---|---|
| Pr | 0.99 | 5.39 | 5.33 | 7.57 | 90.04 | 54.37 | 158° | −1.5 |
| Nd | 0.98 | 5.34 | 5.35 | 7.55 | 90.02 | 53.92 | 156° | −2.5 |
| Sm | 0.96 | 5.25 | 5.41 | 7.48 | 89.90 | 53.11 | 154° | −3.5 |
|    |      | (−1.4%)† | (−0.5%)† | (−1.1%)† | (−0.05°)† | (−3.0%)† |      |      |
| Eu | 0.95 | 5.20 | 5.44 | 7.44 | 89.80 | 52.61 | 152° | −3.8 |
| Gd | 0.94 | 5.17 | 5.46 | 7.40 | 90.10 | 52.22 | 151° | −3.8 |
| Tb | 0.92 | 5.14 | 5.47 | 7.37 | 90.05 | 51.80 | 149° | −3.5 |
| Dy | 0.91 | 5.11 | 5.47 | 7.34 | 90.02 | 51.29 | 148° | −3.8 |
| Ho | 0.90 | 5.09 | 5.47 | 7.31 | 90.04 | 50.88 | 147° | −2.8 |
|    |      | (−1.8%)‡ | (−0.7%)‡ | (−1.5%)‡ | (−0.04°)‡ | (−3.9%)‡ |      |      |
| Er | 0.89 | 5.07 | 5.47 | 7.29 | 90.04 | 50.54 | 146° | −3.0 |

*Reference[47] †Reference [48]; ‡Reference [49]

To model the orthorhombic polymorph of $R$NiO$_3$ ($R$ = Pr–Er), we constructed a computational supercell containing 80 atoms (16 unit formulae (u.f.)) by making 2 × 2 × 1 replications (along crystallographic $a$, $b$, and $c$ directions) of the standard unit cell taken from Ref. 50. Thereafter, we employed a small distortion to the orthorhombic cell setting $\beta$ = 90.75° to obtain the starting structure for monoclinic polymorph following previous DFT works.[9, 28] Several magnetic orderings are possible in $R$NiO$_3$;[17] we have chosen ferromagnetic (FM) ordering for all the calculations in this work inspired by previous successful investigations of MIT in electron doped nickelates using FM ordering.[43-46] Note, that although we employed FM as a representative system throughout this work, the fundamental understanding of the processes underlying MIT in oxygen-deficient nickelates obtained in this work are applicable for other magnetic orderings as well. For instance, all vacancy-ordered square planar $R$NiO$_{2.5}$ compounds are insulating with a band gap of ~0.9 – 1 eV with antiferromagnetic AFM-A ordering (Figure S1); note magnetic ordering controls the extent of band gap opening. Periodic boundary conditions are employed along all directions. The plane wave energy cut off is set at 520 eV. A Γ-centered 3 × 6 × 4 $k$-point grid is used to sample the Brillouin zone. For both orthorhombic and monoclinic $R$NiO$_3$, we optimized the lattice parameters and atomic positions using conjugate gradient algorithm until the atomic forces are less than 1 meV/Å. We employed the tetrahedral method with Blöch corrections to compute the density of states, and at least 20 points between successive high symmetry $k$-points along the $k$-path for band structure calculations. Born effective charges are obtained from density functional perturbation theory.[51, 52]

## III. RESULTS AND DISCUSSION

### A. Atomic structure and thermodynamic stability

We optimized the atomic structure of a series of rare-earth nickelates, $R$NiO$_3$ ($R$ = Pr–Er) within the framework of DFT + $U$ for both possible polymorphs (i.e., orthorhombic (space group: $Pbnm$) and monoclinic ($P2_1/n$)) with ferromagnetic ordering (Table 1, Table S1). Notably, the DFT-predicted structural parameters for both polymorphs are in excellent agreement with previous experiments, with deviations in lattice parameters < 1.9%, unit cell volume < 4%, and monoclinic angle < 0.05°.[19, 43, 48, 49] (Table 1, Table S1). Previously, several DFT and experimental studies have reported that structural distortions in the perovskite lattice are more pronounced with smaller rare-earth cation.[17, 20, 43, 53, 54] Indeed, we found that the average inter-octahedral Ni–O–Ni angle (or tilt angle) deviates further away from the ideal value (i.e., 180° in cubic perovskite) as the ionic radius of the rare-earth cation ($r_R$) decreases, with values ranging from 158° (in PrNiO$_3$) to 146° (in ErNiO$_3$) in monoclinic polymorph (Table 1). Similar trends in inter-octahedral Ni–O–Ni angle are also observed for orthorhombic phase (Table S1). Furthermore, as $r_R$ decreases, the lattice parameters decrease along the crystallographic $a$ and $c$ directions with a concomitant increase in cell length along the $b$-direction, consistent with previous DFT works.[43, 45, 46] Importantly, the monoclinic phase is always slightly more stable as compared to the orthorhombic phase (energetic difference between monoclinic and orthorhombic phases ΔE$_{M-O}$~1.5 – 3.8 meV/u.f.), regardless of the rare-earth cation (Table 1). This stabilization is largely owed to the dominant antiferrodistortive motions involving rotation of $NiO_6$ octahedra in the monoclinic phase.[17]

**Table 2. Structural parameters of vacancy-ordered $RNiO_{2.5}$.** Optimized structural parameters of vacancy-ordered brownmillerite and square planar phases of $RNiO_{2.5}$ with FM ordering obtained from DFT + $U$ calculations. The lengths of unit cell vectors, namely $a$, $b$, and $c$ (listed in Å), and volume $V$ (listed in Å$^3$/u.f.) of a standard unit cell of $RNiO_{2.5}$ are provided for nine different rare-earth cations $R$ = Pr–Er. The monoclinic angle for $RNiO_{2.5}$ are identical to that for the corresponding pristine $RNiO_3$ listed in Table 1. The difference in cohesive energy of brownmillerite and square-planar phases $\Delta E$ are also provided in units of meV/u.f.

| Phase | | Rare Earth cation $R$ in vacancy ordered $RNiO_{2.5}$ | | | | | | | | |
|---|---|---|---|---|---|---|---|---|---|---|
| | | Pr | Nd | Sm | Eu | Gd | Tb | Dy | Ho | Er |
| Brownmillerite | $a$ | 5.16 | 5.36 | 5.13 | 5.09 | 5.21 | 5.05 | 5.06 | 5.07 | 5.11 |
| | $b$ | 5.63 | 5.26 | 5.53 | 5.56 | 5.56 | 5.73 | 5.66 | 5.74 | 5.63 |
| | $c$ | 7.69 | 7.90 | 7.88 | 7.85 | 7.36 | 7.56 | 7.60 | 7.41 | 7.49 |
| | $V$ | 55.79 | 55.67 | 55.94 | 55.55 | 53.31 | 54.63 | 54.33 | 53.99 | 53.81 |
| | $\Delta E$ | 0.41 | 0.42 | 0.43 | 0.44 | 0.46 | 0.47 | 0.47 | 0.47 | 0.47 |
| Square Planar | $a$ | 5.35 | 5.30 | 5.23 | 5.20 | 5.17 | 5.15 | 5.13 | 5.11 | 5.10 |
| | $b$ | 5.52 | 5.53 | 5.54 | 5.53 | 5.53 | 5.52 | 5.51 | 5.50 | 5.50 |
| | $c$ | 7.42 | 7.39 | 7.33 | 7.29 | 7.25 | 7.22 | 7.19 | 7.16 | 7.14 |
| | $V$ | 54.81 | 54.18 | 53.00 | 52.39 | 51.79 | 51.31 | 50.85 | 50.42 | 50.00 |

On the other hand, the square planar $RNiO_{2.5}$ exhibit much more subtle changes in the volume with respect to the pristine phase, with slight expansion for $PrNiO_{2.5}$ (~0.82%) and $NdNiO_{2.5}$ (~0.46%), followed by volumetric compression thereafter with strains gradually rising from $SmNiO_{2.5}$ (~0.27%) to $ErNiO_{2.5}$ (~1%). This behavior is largely owed to the presence of chains of alternating $NiO_6$ octahedra and $NiO_4$ square planar shapes along crystallographic $a$ and $b$ directions. Such an arrangement results in significant reduction of cell length in $c$ direction (~2%) due to availability of empty space for tilting of $NiO_6$ octahedra and $NiO_4$ square planar shapes, irrespective of the rare earth cation. This collapse along the $c$ direction more than compensates the slight expansion in the $ab$ plane (Tables 1, 2). Importantly, the lower volumetric strain significantly enhances the energetic stability of the square planar phase as compared to the brownmillerite. In fact, the cohesive energy of square planar polymorph is at least ~0.41eV/u.f. lower than the brownmillerite phase for any $R$NiO$_{2.5}$ (Table 2).

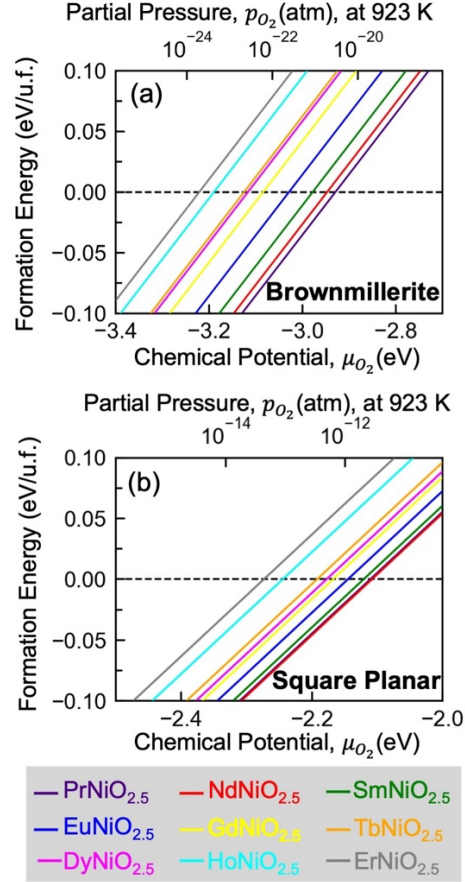

**Figure 2. Thermodynamic stability of vacancy-ordered $R$NiO$_{2.5}$.** Formation energy of $R$NiO$_{2.5}$ ($R$ = Pr–Er) with FM ordering as a function of chemical potential of oxygen for (a) brownmillerite, and (b) square-planar polymorphs. The corresponding partial pressure of oxygen at 923 K are provided in the top axes.

To gain a fundamental understanding of the thermodynamic stability of the vacancy-ordered $R$NiO$_{2.5}$ phases, we evaluated their formation energy $E_f$, which can be written as:[55-57]

$$E_f = E_{R\text{NiO}_{2.5}} - E_{R\text{NiO}_3} + \frac{1}{4}E_{O_2} + \frac{1}{2}\mu_{O_2},$$

where $E_{R\text{NiO}_{2.5}}$, $E_{R\text{NiO}_3}$, and $E_{O_2}$ are the total energy (per unit formula) of $R$NiO$_{2.5}$, $R$NiO$_3$, and O$_2$ molecule, respectively, obtained from DFT + $U$ calculations; while $\mu_{O_2}$ denotes the chemical potential of oxygen. Note, the entropic contributions to formation energy are neglected following previous works on transition metal oxides.[58] In the oxygen-rich limit (i.e., $\mu_{O_2}= 0$), both polymorphs of vacancy-ordered $R$NiO$_{2.5}$ (i.e., brownmillerite and square-planar) exhibit positive formation energies (Table S2), denoting their lower stability as compared to the pristine counterpart $R$NiO$_3$. The formation energy of both these vacancy-ordered phases increases gradually with decreasing size of rare-earth cations (Brownmillerite: 1.46 eV/u.f. (PrNiO$_{2.5}$) – 1.61 eV/u.f. (ErNiO$_{2.5}$); Square Planar: 1.05 eV/u.f. (PrNiO$_{2.5}$) – 1.14 eV/u.f. (ErNiO$_{2.5}$); Table S2). Recent DFT works have reported that oxygen vacancy formation energy in ABO$_3$ perovskites shows a strong direct correlation with the strength of the covalent interactions between the B-site transition metal and oxygen.[59, 60] Following these works, we employ crystal orbital Hamilton population (COHP) analysis,[61, 62] implemented in LOBSTER[63] package, to assess the effect of rare-earth cation on the strength of Ni–O covalent interactions in pristine monoclinic $R$NiO$_3$ phase (relaxed at DFT+ $U$ level) with structural parameters provided in Table 1. This COHP technique offers a route to capture the chemistry of bonding/anti-bonding interactions in the solid state.[61, 62] Notably, COHP is a product of the electronic density of states (DOS) and the overlap Hamiltonian element, which has been demonstrated to provide a better representation of hybridization as compared to the classical crystal orbital overlap population (COOP).[64] Furthermore, integrated crystal orbital Hamilton population (ICOHP) gives a qualitative measure of the strength of the covalent interactions; specifically, more negative ICOHP denotes stronger covalent interaction.[59-63] The calculated mean ICOHP value for Ni–O bond within the NiO$_6$ octahedra becomes more negative as the size of rare-earth cation decreases (ICOHP for PrNiO$_3$: –0.36 eV; and ErNiO$_3$: –0.40 eV). Evidently, the Ni–O covalent interactions are stronger in nickelates with smaller rare-earth cations; this in turn, leads to a higher formation energy for vacancy-ordered nickelates with smaller rare-earth ions.

The chemical potential of oxygen $\mu_{O_2}$ can be related to the temperature and oxygen partial pressure in the environment, providing a route to understand thermodynamic stability of vacancy-ordered phases under various synthesis conditions. Assuming that oxygen reservoir behaves as an ideal gas, we can determine the chemical potential of oxygen for a given pressure $P$, and temperature $T$ using the following relationship:[55-57]

$$\mu_{O_2}(T,P) = \mu_{O_2}(T,P_0) + \frac{1}{2} k_b T \ln\left(\frac{P}{P_0}\right),$$

where $P_0$ is the ambient pressure, and $k_b$ is the Boltzmann constant. Using typical synthesis temperature of 650 °C, [58] we took $\mu_{O_2}(T,P_0) = 1.05$ eV from thermodynamic tables.[65] Figure 3 shows the formation energy of both polymorphs of $R$NiO$_{2.5}$ (i.e., brownmillerite and square planar) as a function of chemical potential of oxygen $\mu_{O_2}$ for the entire series of rare earth cations ($R$ = Pr–Er). Note that for all values of $\mu_{O_2}$ (and corresponding partial pressure of oxygen) that yield positive formation energies for a given vacancy-ordered nickelate, the corresponding pristine $R$NiO$_3$ (monoclinic) is more stable. As expected from the trends in formation energy at the oxygen-rich limit, we find that for both brownmillerite and square planar phases, the critical oxygen partial pressure required to spontaneously form $R$NiO$_{2.5}$ progressively reduces with decreasing size of the rare-earth cation. For instance, brownmillerite PrNiO$_{2.5}$ is stable at $p_{O_2}< 3 \times 10^{-21}$ atm, while ErNiO$_{2.5}$ requires more oxygen-lean environment with $p_{O_2}< 2 \times 10^{-24}$ atm. Similarly, the critical oxygen partial pressure required for forming square planar PrNiO$_{2.5}$ and ErNiO$_{2.5}$ are ~$3 \times 10^{-12}$ atm and ~$4 \times 10^{-14}$ atm, respectively. Note, such partial pressures are certainly be realized in the laboratory. Furthermore, we find that for a given $R$NiO$_{2.5}$ at any temperature, formation of brownmillerite phase requires much leaner oxygen environment (i.e., several orders of magnitude lower $p_{O_2}$) as compared to the square-planar counterparts.

## B. Electronic Structure

Next, we elucidate the effect of spatially ordered OVs on the electronic structure of rare-earth nickelates using PrNiO$_3$ as a representative case. Figure 3 compares the electronic structure of Ni atoms in pristine monoclinic PrNiO$_3$ with those in square planar PrNiO$_{2.5}$ (i.e., the most stable vacancy-ordered phase at this stoichiometry). The *3d* orbitals of Ni undergo different crystal field splitting depending on coordination of oxygen ligands around them. For Ni centered at octahedral NiO$_6$, the *3d* orbitals split into two distinct energy levels, namely $t_{2g}$ (composed of $d_{xy}, d_{yz}$, and $d_{xz}$ orbitals) at lower level, and $e_g$ (composed of $d_{x^2-y^2}$ and $d_{z^2}$ orbitals) at the higher level (Figure 3(a)). On the other hand, for a square planar coordination (such as in NiO$_4$), the crystal field splitting gives rise to four levels: $(d_{yz}, d_{xz})$, $d_{xy}$, $d_{z^2}$, and $d_{x^2-y^2}$, in increasing order of energy (Figure 3(a)). Pristine PrNiO$_3$ is metallic, as indicated by the DOS in Figure 3(b). Here, all the Ni atoms possess an octahedral coordination with O ligands, and are in 3+ valence state. This yields a high spin state for all Ni with electronic configuration $t_{2g}^6 e_g^1$, and a magnetic moment of ~$0.9\mu_B$ (Figure 3(a)). Introduction of 0.5 OV per Ni atom to obtain the square-planar PrNiO$_{2.5}$ phase is akin to doping the system with $1e^-$/Ni. These electrons localize into the unoccupied $e_g$ spin-up state in the bottom of the conduction band in PrNiO$_3$ and move them to the top of the valence band by shifting the Fermi level (see region highlighted by dotted squares in Figures 3(b, c)) similar to previous DFT reports; and reduce the octahedral $Ni^{3+}$ to high spin $Ni^{2+}$ (magnetic moment ~1.5$\mu_B$). We also confirmed electron filling of octahedral $Ni^{2+}$ using Wannier analysis[66] (Figure S2). Note, such electron localization have also been observed upon electron doping with addition of protons, lithium, and other cations.[4, 9, 43-46, 67] Such electron filling increases the energy of the unoccupied states due to on-site electron correlations opening up an on-site Mott gap of ~0.2 eV between the occupied spin up $e_g$ and spin-down $e_g$ states of octahedral $Ni^{2+}$ (Figure 3(c)). The square planar Ni are also all in 2+ valence state; the crystal field splitting for these Ni yields low spin $Ni^{2+}$ (~0.2$\mu_B$) with fully filled $d_{yz}$, $d_{xz}$, $d_{xy}$, and $d_{z^2}$ orbitals (in the valence band) and empty $d_{x^2-y^2}$ states (in the conduction band) (Figure 3(c)). This results in the crystal field splitting gap of ~1.1 eV on the square-planar $Ni^{2+}$ between the occupied deep valence $d_{z^2}$ states and unoccupied $d_{x^2-y^2}$ states at the bottom of the conduction band (Figure 3(c)). However, the relative energies of the octahedral/square planar $Ni^{2+}$ orbitals leads to (a) conduction band minimum (CBM) with nearly equal contributions from unoccupied spin-down $e_g$ states of octahedral $Ni^{2+}$ and spin-down $d_{x^2-y^2}$ states of square planar $Ni^{2+}$, and (b) valence band maximum (VBM) consisting of occupied spin up $e_g$ states of octahedral $Ni^{2+}$ alone; yielding a band gap of ~0.2 eV. This finding is slightly different from previous DFT + $U$ calculations using standard PBE,[29] which show that the conduction band edge is dictated by the unoccupied $d_{x^2-y^2}$ states of square planar $Ni^{2+}$ alone. Note, we did not observe any separation of the unoccupied octahedral $e_g$ states and square planar $d_{x^2-y^2}$ states even upon increasing $U$ (up to 4.6 eV). On the other hand, both standard PBE (Ref. 29) and PBEsol (this work) indicate that the valence band maximum is governed by the electron-filled $e_g$ state of octahedral $Ni^{2+}$. Nevertheless, the electronic structure of $R$NiO$_{2.5}$ is decided by the combined effect of the on-site Coulomb repulsion in high spin $Ni^{2+}$ and crystal field splitting in different NiO$_x$ polyhedra, consistent with PBE + $U$.[29]

The significance of crystal field splitting for band gap opening is more evident in the case of brownmillerite $R$NiO$_{2.5}$ (Figure S3). Similar to square-planar phase, electron filling leads to an on-site Mott gap between the unoccupied spin down and occupied spin-up $e_g$ state (at the Fermi level) of high spin (~1.5$\mu_B$) octahedral $Ni^{2+}$. However, the crystal field splitting of tetrahedrally coordinated Ni (in NiO$_4$) inverts the energetic ordering of $e_g$ and $t_{2g}$, which leads to a high spin $Ni^{2+}$ (~1.5$\mu_B$). Interestingly, although this opens up an on-site gap of ~1 eV between the spin-down unoccupied and spin-up occupied $t_{2g}$ states on $Ni^{2+}$ tetrahedra, the bottom of the unoccupied states is at the Fermi level, and yields a metallic state for all brownmillerite $R$NiO$_{2.5}$ (Figure S3). Since the brownmillerite $R$NiO$_{2.5}$ is energetically less stable than the counterpart square-planar phase (Table 2, Figure 2) and does not feature a band gap (Figure S3), the next section focuses on square-planar $R$NiO$_{2.5}$ alone.

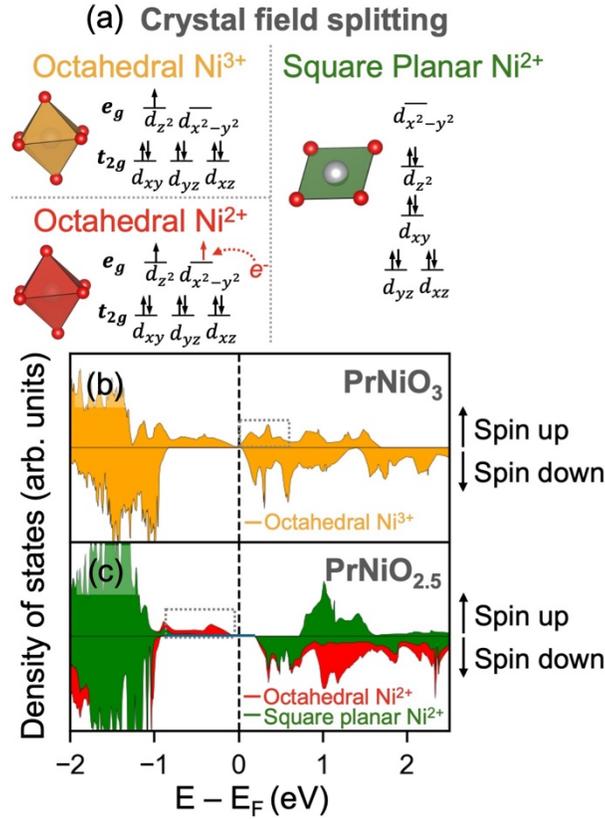

**Figure 3. Electronic structure of vacancy-ordered $R$NiO$_{2.5}$ in square-planar polymorph.** (a) Schematic illustration of crystal field splitting of Ni *3d* orbitals in octahedral and square planar coordination with oxygen ligands. The electronic configuration for each coordination is shown for different oxidation states of Ni. Site projected electronic density of states of Ni *3d* orbitals for (b) Ni$^{3+}$ (all centered at NiO$_6$ octahedra) in pristine PrNiO$_3$ (orange), are compared with those for (c) Ni$^{2+}$ belonging to NiO$_6$ octahedra (red), and square planar NiO$_4$ polyhedra (green) in square-planar PrNiO$_{2.5}$. We considered FM ordering for both PrNiO$_3$ and PrNiO$_{2.5}$. The vertical dotted line denotes the Fermi level. The states involved in localization of electron in the Ni centered at NiO$_6$ centered are highlighted by the dotted rectangle.

## C. Modulation of Electronic Band Gap by Rare-Earth Cation

Square-planar $R$NiO$_{2.5}$ opens up a band gap owing to the combined effect of crystal field splitting in NiO$_6$/NiO$_4$ polyhedra and on-site Coulomb repulsion in high-spin Ni$^{2+}$, irrespective of the size of the rare-earth cation (Figures 4, 5, Figure S4). For all the rare-earth cations, the valence band maximum (VBM) is composed of the $e_g$ state of Ni belonging to NiO$_6$ octahedra on whom electrons from OV localize (Figure 4, Figure S4). Also, similar to PrNiO$_{2.5}$, the conduction band minimum (CBM) has nearly equal contributions from unfilled (a) $e_g$ state of Ni belonging to NiO$_6$ octahedra, and (b) $d_{x^2-y^2}$ state of low-spin Ni$^{2+}$ present in the square-planar shape for all $R$NiO$_{2.5}$ (Figure 4). In all cases, both VBM and CBM lie at the Γ-point, yielding a direct band gap (Figure 4) consistent with the previous PBE + $U$ calculations on SmNiO$_{2.5}$.[29]

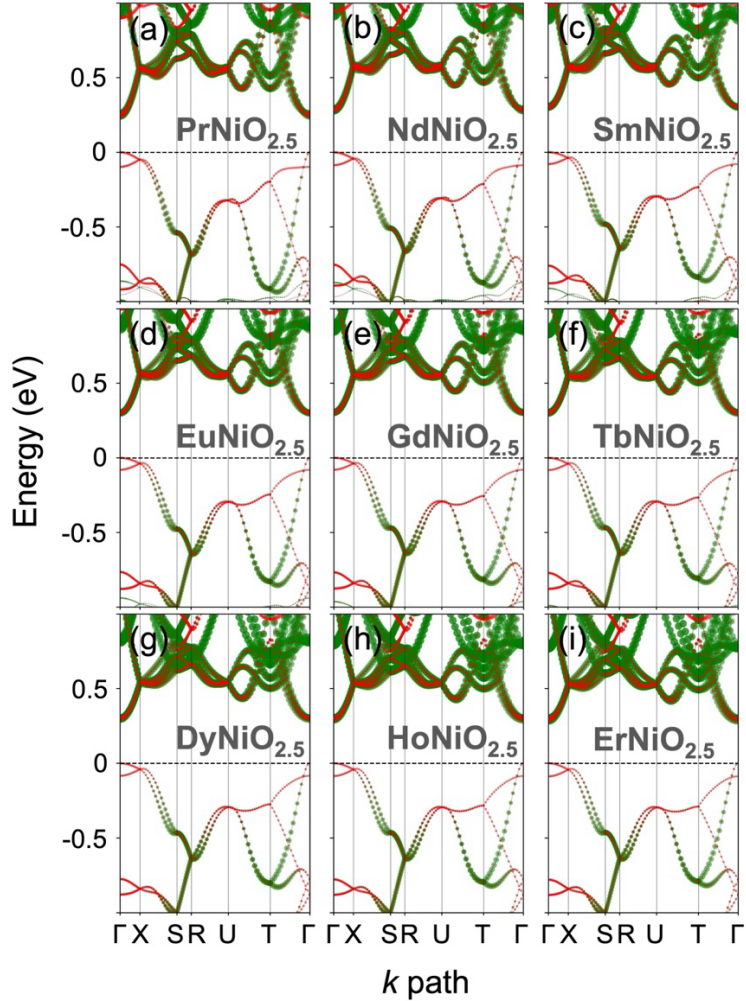

**Figure 4. Electronic band structure of square-planar $R$NiO$_{2.5}$.** Band structure for square-planar $R$NiO$_{2.5}$ with FM ordering for various rare-earth cations, namely (a) Pr, (b) Nd, (c) Sm, (d) Eu, (e) Gd, (f) Tb, (g) Dy, (h) Ho, and (i) Er (listed in decreasing order of their ionic radius) showing the orbital character of Ni $3d$ belonging to NiO$_6$ octahedra (red), and NiO$_4$ square plane polyhedra (green). The Fermi level is set at 0 eV.

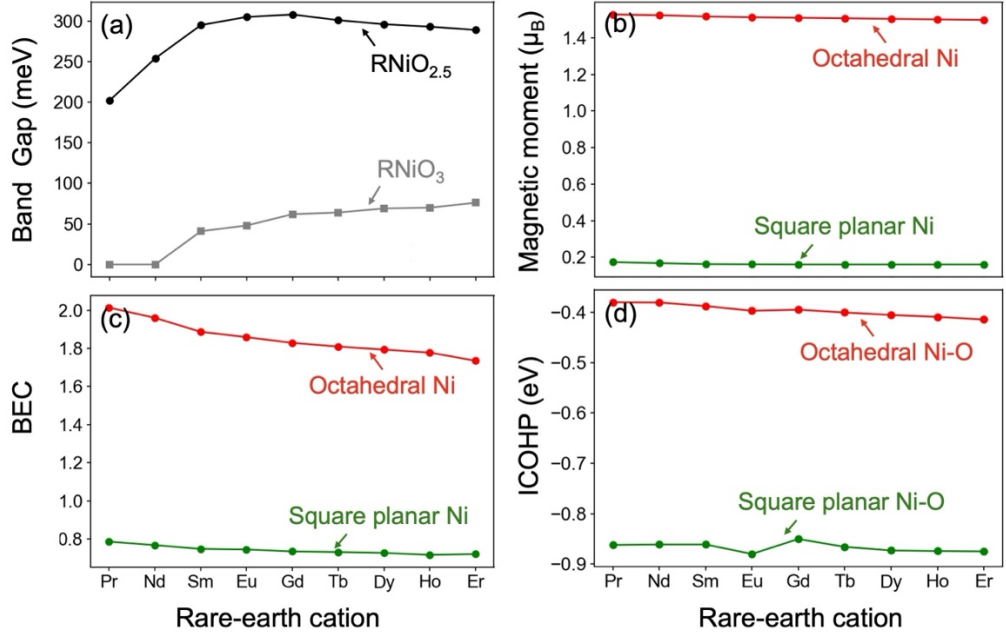

**Figure 5. Modulation of electronic structure of $R$NiO$_{2.5}$ using rare-earth cation.** (a) Band gap of square-planar $R$NiO$_{2.5}$ with FM ordering compared against pristine counterparts for various rare-earth cations $R$ = Pr–Er. We have also shown (b) Born Effective charge of Ni (c) Magnetic moment of Ni, and (c) ICOHP of Ni–O bond belonging to NiO$_6$ octahedra and NiO$_4$ square planar shapes in square-planar $R$NiO$_{2.5}$ for the entire series of rare-earth cations.

Figure 5(a) shows compares the band gap of square-planar $R$NiO$_{2.5}$ with that for pristine $R$NiO$_3$ as a function of size of rare-earth cation. PrNiO$_3$ and NdNiO$_3$ are metallic; while all remaining $R$NiO$_3$ (R = Sm–Er) have a small band gap (~41–76 meV) due to a slight bond disproportionation facilitated by lattice distortions involving tilting of NiO$_6$ octahedra.[17] For all rare-earth cations, the band gap in square-planar $R$NiO$_{2.5}$ is significantly larger (~200–250 meV) than their pristine counterparts (Figure 5(a)). Importantly, the on-site gap on the low-spin square planar Ni$^{2+}$ arising from crystal field splitting (i.e., between the occupied $d_{z^2}$ and unoccupied $d_{x^2-y^2}$ levels) remains largely constant (~1.1 eV) over the entire series of rare-earth cations (Figure S4). In effect, the modulation in the band gap of $R$NiO$_{2.5}$ induced by $R$ is primarily dictated by the on-site Coulomb repulsion within the $e_g$ states of Ni centered at NiO$_6$ octahedra (Figure S4). As the size of rare-earth cation decreases, the band gap in $R$NiO$_{2.5}$ increases significantly from Pr (202 meV) to Sm (300 meV). This increase can be ascribed to the diminishing overlap between the unoccupied $e_g$ states of Ni in NiO$_6$ octahedra and O-$2p$ states due to significant drop in inter-octahedral Ni–O–Ni angle from PrNiO$_{2.5}$ (~151º) to SmNiO$_{2.5}$ (~146º); which raises the on-site Coulomb repulsion.[17,]

[28, 29] As the $r_R$ decreases beyond Sm, band gap of $R$NiO$_{2.5}$ remains largely constant with values ranging between ~290–300 meV for $R$ = Eu–Er, owing to limited changes in the inter-octahedral Ni–O–Ni angle.

As aforementioned, the physical origin of the band gap in $R$NiO$_{2.5}$ is identical to those identified in Figure 3 for all rare-earth cations (Figure 4, Figure S4). All $R$NiO$_{2.5}$ show a high-spin octahedral Ni$^{2+}$ (~1.5 $\mu_B$) that arises from electron localization on NiO$_6$ and low-spin square planar Ni$^{2+}$ (~0.2 $\mu_B$) owing to crystal field splitting (Figure 5(b)). Although the formal oxidation state of Ni atoms belonging to both NiO$_6$ octahedra and NiO$_4$ square plane is 2+ in $R$NiO$_{2.5}$, the Born effective charge (BEC) of only the octahedral Ni is close to 2+ (~1.8–2, with lower values for smaller $R$) (Figure 5(c)). On the other hand, the BEC of square planar Ni are much lower (~0.8). This is mainly due to the strong hybridization of the $3d$ orbitals of the square planar Ni with the $2p$ orbitals of the four oxygen ligands, as indicated by the more negative ICOHP value (~ –0.87 eV) as compared to that for Ni–O bond in NiO$_6$ octahedra (~ –0.38 eV) as shown in Figure 5(d). In essence, the square planar $R$NiO$_{2.5}$ features chains of corner-sharing NiO$_6$ octahedra and NiO$_4$ square planar polyhedra, wherein the Ni–O bonds in NiO$_6$ are predominantly ionic, while those belonging to NiO$_4$ exhibit more covalent character.

## IV. CONCLUSIONS

We employed DFT + U calculations to investigate the structure, energetics, and electronic behavior of oxygen-deficient $R$NiO$_{2.5}$ ($R$ = Pr – Er) in two typical vacancy ordered configurations, namely (a) brownmillerite consisting of alternating planes of NiO$_6$ octahedra and NiO$_4$ tetrahedra and (b) square planar comprising chains of corner-sharing NiO$_6$ octahedra and NiO$_4$ square planar polyhedra. Our predicted formation energies indicate that the square planar polymorph is always more stable (~0.4 eV/u.f.) than the brownmillerite counterpart for all $R$NiO$_{2.5}$. Furthermore, our COHP analysis shows that smaller $R^{3+}$ cations enhance the strength of the Ni-O covalent interactions in pristine $R$NiO$_3$, which in turn, increases the formation energy for $R$NiO$_{2.5}$ for smaller $R$. This translates to a gradual decrease in critical partial pressure of oxygen required for spontaneous formation of $R$NiO$_{2.5}$ from $R$NiO$_3$ with increase in ionic radius of $R$. In terms of the electronic properties of $R$NiO$_{2.5}$, we find that electrons from OVs localize on the NiO$_6$ octahedra, reducing the Ni to a 2+ state from a 3+ state, yielding an high-spin Ni$^{2+}$ electronic configuration of $e_g^2$. The on-site Coulomb repulsion in the high spin Ni$^{2+}$ opens up a Mott-type gap. At the same time, the electronic structure of $R$NiO$_{2.5}$ changes as compared to pristine $R$NiO$_3$ due to the crystal field splitting of tetrahedral NiO$_4$ (brownmillerite) and NiO$_4$ square planar (square planar) configurations. All the brownmillerite $R$NiO$_{2.5}$ turn out to be metallic due to the inversion of the crystal field splitting in NiO$_4$ tetrahedra (as compared to NiO$_6$ octahedra in pristine $R$NiO$_3$), which places the unoccupied $t_{2g}$ states of NiO$_4$ tetrahedra at the Fermi level. On the other hand, all the square planar $R$NiO$_{2.5}$ show a direct band gap ~0.2 – 0.3 eV at the Γ-point, depending on the $R^{3+}$ cation. For all square planar $R$NiO$_{2.5}$, the VBM comprises of the filled $e_g$ state of Ni belonging to NiO$_6$ octahedra, while the CBM features nearly equal contribution from the unoccupied $e_g$ state of octahedral Ni, and $d_{x^2-y^2}$ states of square-planar Ni. Finally, our predicted ICOHP values indicate that the Ni–O bond belonging to NiO$_6$ octahedra has a much higher ionic character as compared to that in NiO$_4$ square planar polyhedra in square planar $R$NiO$_{2.5}$. The findings presented here are crucial for advances in design of quantum materials for a wide range of applications in emerging areas of brain-like computing platforms, and novel memory devices.

## ACKNOWLEDGEMENTS

This work was supported by the U.S. Department of Energy, Office of Science, Basic Energy Sciences, under Award # DE-SC0021229. We also acknowledge the resources of the National Energy Research Scientific Computing Center, a DOE Office of Science User Facility supported by the Office of Science of the U.S. Department of Energy under Contract No. DE-AC02-05CH11231.

## DATA AVAILABILITY STATEMENT

The data that supports the findings of this study are available within the article and its supplementary material.

## CONFLICT OF INTEREST

The authors have no conflicts to disclose

## AUTHOR CONTRIBUTIONS

B. Narayanan conceived the idea and supervised the research. D. Bhagat performed all the DFT calculations, and analyzed the data with R. Pidathala. D. Bhagat and B. Narayanan wrote the manuscript with inputs from R. Pidathala. All authors discussed the results and revised the manuscript.

# First-principles investigation of electronic transitions in vacancy-ordered rare-earth perovskite nickelates

Devang Bhagat, Ranga Teja Pidathala and Badri Narayanan[*]

Department of Mechanical Engineering, University of Louisville, Louisville KY 40223, USA

[*]Corresponding author, Email: badri.narayanan@louisville.edu

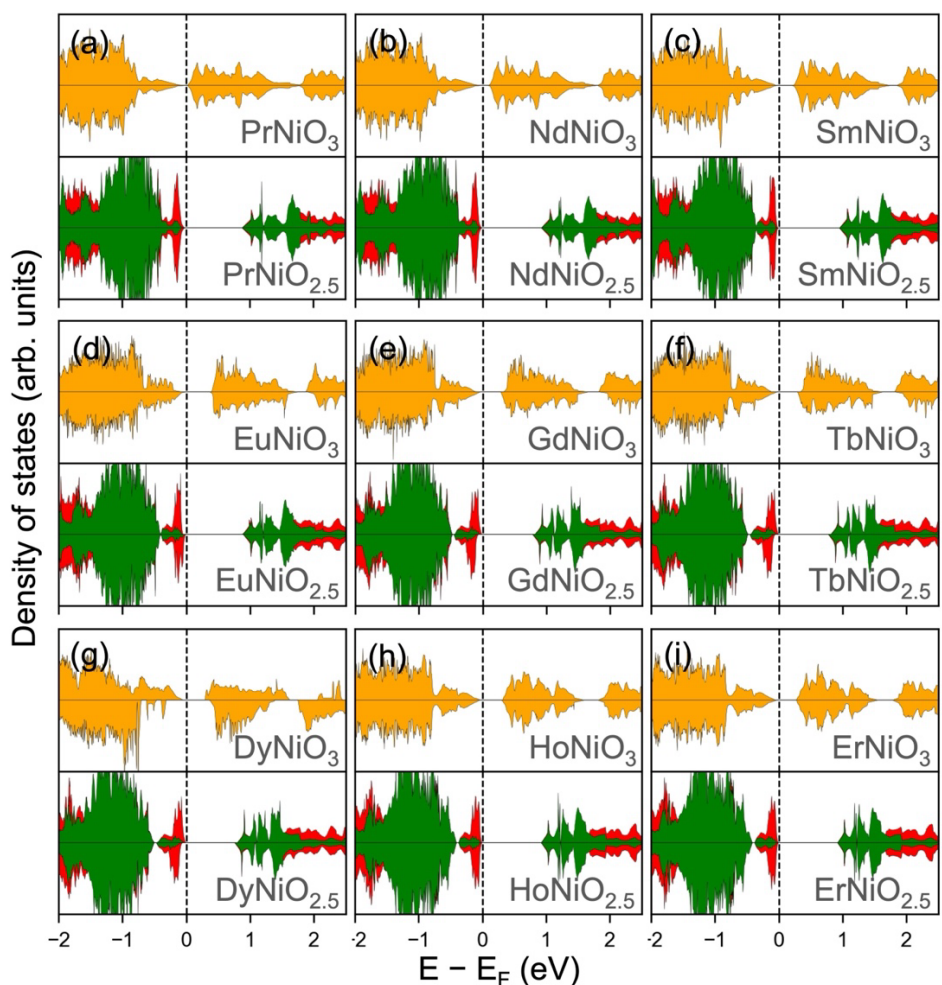

**Figure S1. Effect of oxygen vacancies on electronic structure of square planar $R$NiO$_{2.5}$ with antiferromagnetic AFM-A ordering.** Site projected electronic density of states of Ni *3d* orbitals for Ni$^{3+}$ in NiO$_6$ octahedra in pristine monoclinic $R$NiO$_3$ (orange), as well as Ni$^{2+}$ centered at NiO$_6$ octahedra (red) and square planar (green) polyhedra in square planar phase of $R$NiO$_{2.5}$ for the entire series of rare-earth cations, namely (a) Pr, (b) Nd, (c) Sm, (d), Eu, (e) Gd, (f) Tb, (g) Dy, (h) Ho, and (i) Er (listed in decreasing order of their ionic radius). The vertical dotted line denotes the Fermi level.

**Table S1. Optimized structural properties of a conventional unit cell of orthorhombic $R$NiO$_3$ (space group: *Pbnm*) with ferromagnetic ordering for various rare-earth cations ($R$ = Pr – Er) obtained from DFT + $U$ calculations.** The lattice parameters, namely *a*, *b*, and *c* are listed in Å, while the unit cell volume and the inter-octahedral Ni–O–Ni tilt angle $\theta$ are provided in Å/u.f. and degrees, respectively. Percentage deviation of the structural properties predicted by DFT + $U$ (this work) from the respective values reported by previous experiments (wherever available) are noted within parenthesis.

|  | Rare Earth Cation $R$ in orthorhombic $R$NiO$_3$ (Space Group: *Pbnm*) | | | | | | | | |
|---|---|---|---|---|---|---|---|---|---|
|  | **Pr** | **Nd** | **Sm** | **Eu** | **Gd** | **Tb** | **Dy** | **Ho** | **Er** |
| *a* | 5.39 (–0.5%)[*] | 5.34 (–0.9%)[*] | 5.25 (–1.5%)[†] | 5.20 (–1.7%)[†] | 5.16 (–1.9%)[†] | 5.13 (–0.1%)[‡] | 5.11 (–1.9%)[†] | 5.09 (–1.8%)[†] | 5.07 (–1.5%)[*] |
| *b* | 5.33 (–1.1%)[*] | 5.34 (–0.8%)[*] | 5.41 (–0.4%)[†] | 5.45 (–0.1%)[†] | 5.47 (–0.3%)[†] | 5.48 (0.15%)[‡] | 5.48 (–0.4%)[†] | 5.48 (–0.5%)[†] | 5.48 (–0.3%)[*] |
| *c* | 7.56 (–0.9%)[*] | 7.54 (–0.9%)[*] | 7.47 (–1.2%)[†] | 7.43 (–1.4%)[†] | 7.39 (–1.6%)[†] | 7.36 (–0.1%)[‡] | 7.33 (–1.6%)[†] | 7.30 (–1.7%)[†] | 7.27 (–1.2%)[*] |
| *V* | 54.30 (–2.3%)[*] | 53.87 (–2.4%)[*] | 53.09 (–3%)[†] | 52.63 (–3.3%)[†] | 52.13 (–3.8%)[†] | 51.70 (–0.2%)[‡] | 51.28 (–3.9%)[†] | 50.87 (–4%)[†] | 50.48 (–2.9%)[*] |
| $\theta$ | 159° | 157° | 153.3° | 151.8° | 150.3° | 149.1° | 148.1° | 147.1° | 146.2° |

[*]Reference 1; [†]Reference 2; [‡] Reference 3

**Table S2. Thermodynamic stability of vacancy-ordered $R$NiO$_{2.5}$ in oxygen-rich environment.** Formation energy of the two polymorphs of vacancy-ordered $R$NiO$_{2.5}$, namely brownmillerite (Figure 1(b)) and square-planar (Figure 1(a)) at the oxygen-rich limit for various rare-earth cations ($R$ = Pr–Er). All the formation energy values are provided in eV/u.f.

|  | Rare Earth cation $R$ in vacancy-ordered $R$NiO$_{2.5}$ | | | | | | | | |
|---|---|---|---|---|---|---|---|---|---|
|  | **Pr** | **Nd** | **Sm** | **Eu** | **Gd** | **Tb** | **Dy** | **Ho** | **Er** |
| Brownmillerite | 1.46 | 1.47 | 1.49 | 1.51 | 1.54 | 1.56 | 1.56 | 1.59 | 1.61 |
| Square Planar | 1.05 | 1.05 | 1.06 | 1.07 | 1.08 | 1.09 | 1.09 | 1.12 | 1.14 |

| (a) Ni $d_{x^2-y^2}$ (spin up) | (b) Ni $d_{z^2}$ (spin up) |
|---|---|
| 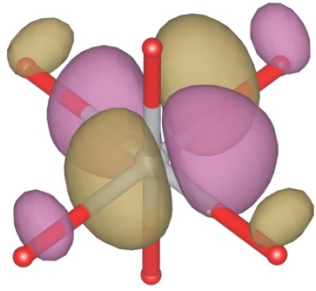 | 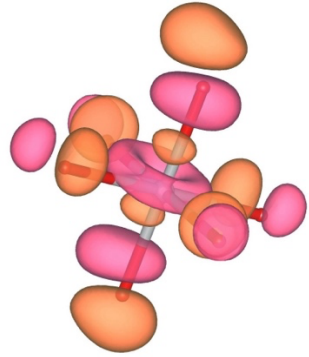 |
| (c) Ni $d_{x^2-y^2}$ (spin down) | (d) Ni $d_{z^2}$ (spin down) |
| 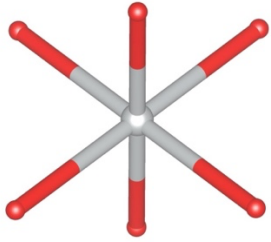 | 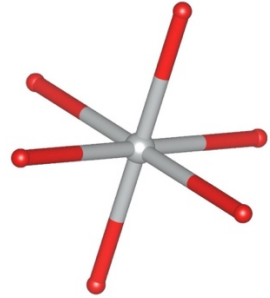 |

**Figure S2. Occupancy of $e_g$ states in Ni belonging to NiO$_6$ octahedra in a representative square-planar RNiO$_{2.5}$ with ferromagnetic ordering.** Maximally located Wannier functions (MLWFs) on Ni centered at a selected NiO$_6$ octahedra in PrNiO$_{2.5}$ for (a) $d_{x^2-y^2}$ (spin up), (b) $d_{z^2}$ (spin up), (c) $d_{x^2-y^2}$ (spin down), and (d) $d_{z^2}$ (spin down) using an isosurface value of 2.5. The MLWFs were obtained by restricting the localization procedure to 15 occupied energy levels (i.e., in the valence band) near the Fermi level in Wannier90 package.[4] We employed three generic $p$ orbitals centered at each O anion, and 5 $d$ orbitals centered at two nearest neighbor cations (one belonging to octahedron and the other to a square planar polyhedron) as the initial orbitals following Ref. 5.

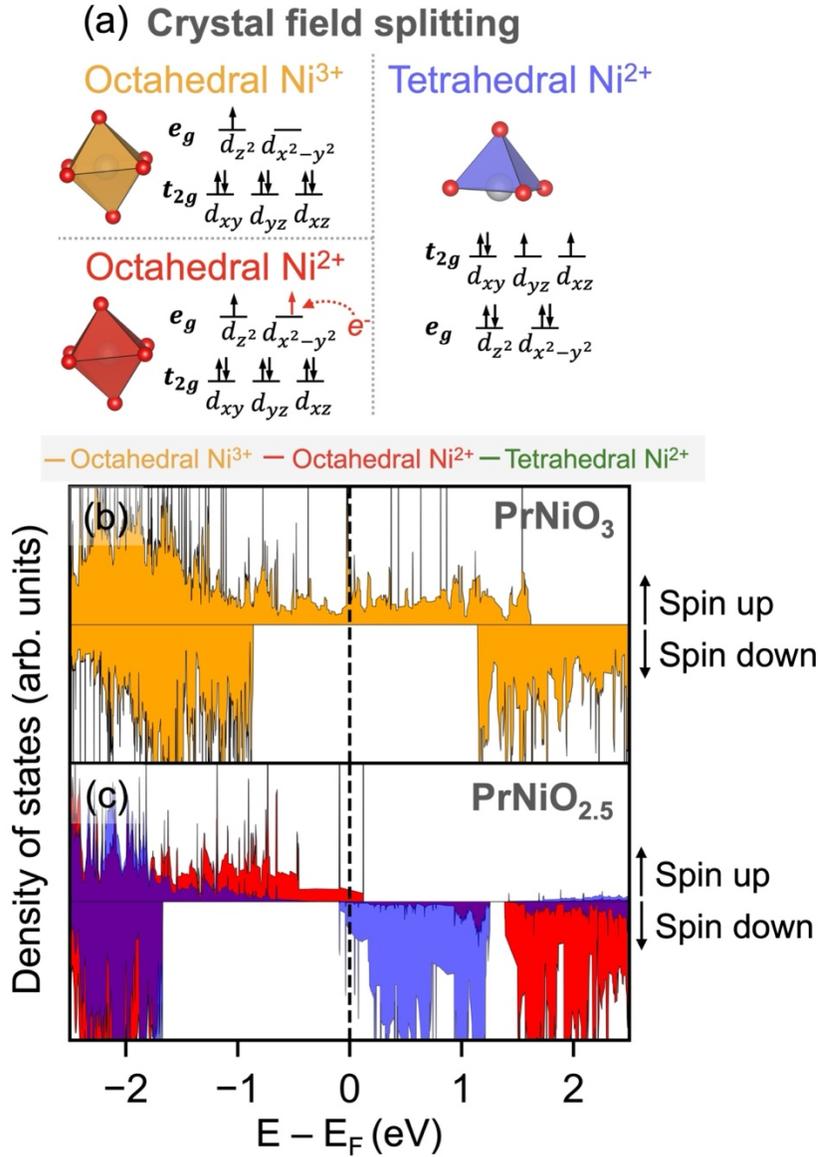

**Figure S3. Electronic structure of vacancy-ordered RNiO$_{2.5}$ in brownmillerite polymorph with ferromagnetic ordering.** (a) Schematic illustration of crystal field splitting of Ni *3d* orbitals in octahedral and tetrahedral coordination with oxygen ligands. The electronic configuration for each coordination is shown for different oxidation states of Ni. Site projected electronic density of states of Ni *3d* orbitals for (b) Ni$^{3+}$ (all centered at NiO$_6$ octahedra) in pristine PrNiO$_3$ (orange), are compared with those for (c) Ni$^{2+}$ belonging to NiO$_6$ octahedra (red) , and square planar NiO$_4$ polyhedra (green) in brownmillerite PrNiO$_{2.5}$. The vertical dotted line denotes the Fermi level.

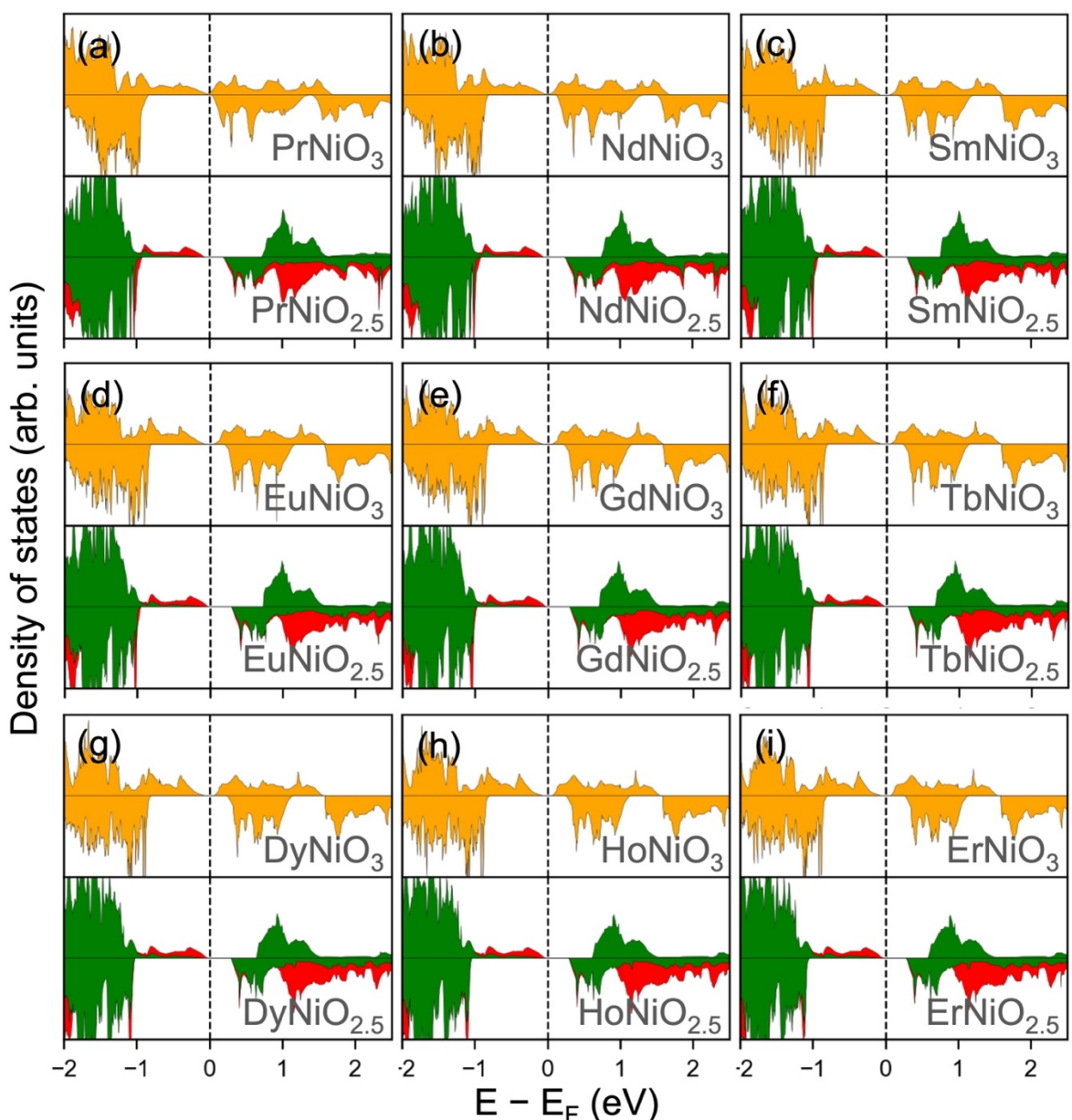

**Figure S4. Electronic structure of square planar $R$NiO$_{2.5}$ with ferromagnetic ordering.** Site projected electronic density of states of Ni $3d$ orbitals for Ni$^{3+}$ in NiO$_6$ octahedra in pristine monoclinic $R$NiO$_3$ (orange), as well as Ni$^{2+}$ centered at NiO$_6$ octahedra (red) and square planar (green) polyhedra in square planar phase of $R$NiO$_{2.5}$ for the entire series of rare-earth cations, namely (a) Pr, (b) Nd, (c) Sm, (d) Eu, (e) Gd, (f) Tb, (g) Dy, (h) Ho, and (i) Er (listed in decreasing order of their ionic radius). The vertical dotted line denotes the Fermi level.